\documentclass[10pt,conference]{IEEEtran}
\usepackage{amsfonts,amssymb,amsmath}
\usepackage{graphicx}
\begin{document}

\title{MDS codes on the erasure-erasure wiretap channel}

\author{
\IEEEauthorblockN{Arunkumar Subramanian, Steven W. McLaughlin }
\IEEEauthorblockA{School of Electrical and Computer Engineering\\
Georgia Institute of Technology\\
Atlanta, GA 30332, USA\\
Email: arunkumar@gatech.edu, swm@ece.gatech.edu}
}

\maketitle

\begin{abstract}
This paper considers the problem of perfectly secure communication on a modified version of Wyner's wiretap channel II where both the main and wiretapper's channels have some erasures. A secret message is to be encoded into $n$ channel symbols and transmitted. The main channel is such that the legitimate receiver receives the transmitted codeword with exactly $n - \nu$ erasures, where the positions of the erasures are random. Additionally, an eavesdropper (wire-tapper) is able to observe the transmitted codeword with $n - \mu$ erasures in a similar fashion. This paper studies the maximum achievable information rate with perfect secrecy on this channel and gives a coding scheme using nested codes that achieves the secrecy capacity.
\end{abstract}

\section{Introduction}
The wire-tap channel was introduced by Wyner \cite{wiretap1}, where a transmitter (Alice) wants to convey a secret message to a legitimate receiver (Bob) through a discrete memoryless channel (DMC). The message must be kept secret from an eavesdropper (Eve) who has a degraded version of the legitimate receiver's observation. Wyner has studied the information rates at which complete secrecy is possible in this system. This work was furthered by that of Csiszar and Korner \cite{1055892}, who generalized the secrecy concept to general wire-tap channels, where the two receivers have noisy observations of the same channel transmission. They have studied the maximum possible secret information rate for this generalized wire-tap channel.

The wire-tap channel II was studied in \cite{wiretap2}, where the transmission length is fixed to $n$. Alice must convey a $k$ symbol message to Bob by transmitting $n$ symbols over the channel. Bob receives these symbols without noise and Eve can observe a fixed number, $\mu$, of the transmitted symbols. Ozarow and Wyner provided a stochastic coding scheme based on cosets of linear codes for this channel. This scheme ensured successful decoding by Bob while Eve is kept completely ignorant as long as a good linear code is chosen and $\mu$ is not too large.

In this paper, we study a modified version of wire-tap channel II where the main channel can also have a fixed number of erasures, $n - \nu$. In other words, Bob can observe any $\nu$ of the $n$ channel symbols, and Eve can observe any $\mu$ of the channel symbols. Alice must devise a coding scheme that guarantees successful decoding by Bob while Eve can obtain no information about the message. We prove that the maximum amount of secret information that can be conveyed in this channel is $\nu - \mu$, assuming $\nu \geq \mu$.   We then show how a nested coding scheme (\cite{1003821}) can be used to achieve the secrecy capacity.

A simple solution to combat erasures on the main channel is to use the existing Ozarow-Wyner coding scheme \cite{wiretap2} (described shortly) , with an outer code for error correction. We propose a coding scheme based on Ozarow-Wyner's coset coding which is equivalent to adding an outer code. We use two codes $C, C^*$ that have only the zero codeword as the common element. We encode the secret message using $C$ and encode a random vector with $C^*$, and transmit the sum of the two resultant codewords. This formulation is easier to analyze compared to the cascaded encoder formulation since the inner and outer codes are combined into a single encoder. Also, it can be noted that when $C + C^*$ and $C^*$ are MDS codes, then the information rate equals the secrecy capacity of this channel. When $\nu - \mu = 1$, this coding scheme with MDS codes becomes identical to the coding scheme used by Shamir \cite{359176} for the $(k, n)$-threshold secret sharing scheme. 

To study the performance of our coding technique, we use the Dimension-Length Profile (DLP) property of linear block codes. The basic idea behind the DLP and its relation to wire-tap channels was first published by Wei \cite{133259}. The DLP property and its related LDP property were rigorously defined and studied by Forney \cite{340452}. Mitrpant, et al. \cite{1397965} studied the secrecy capacity of wiretap channel II under a special case where some of the information bits are revealed. Their analysis uses the DLP properties of linear block codes and their expressions for secrecy capacity are similar to those in our paper. 

\begin{figure}
	\centering
		\includegraphics[width=0.48\textwidth]{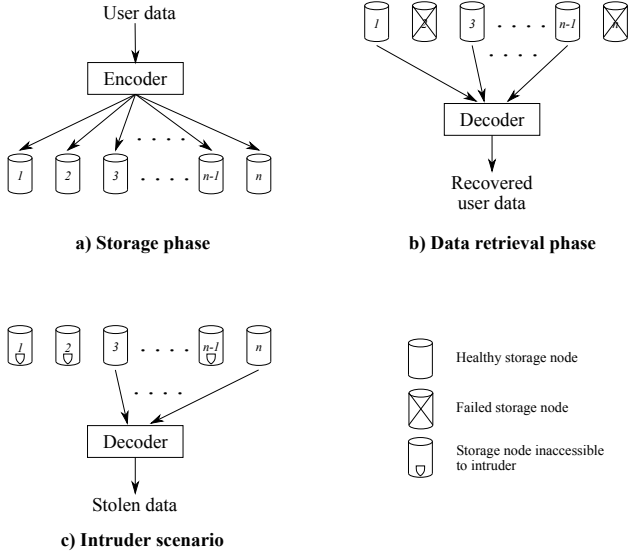}
		\caption{Secure distributed storage system with an eavesdropper}
		\label{fig:storage_example}
\end{figure}

A practical example of this modified wire-tap channel II problem is in distributed storage depicted in figure \ref{fig:storage_example}. A user wants to store a secret information in $n$ data nodes. These data nodes are susceptible to random failure. We want to design a system that can reconstruct the secret data from any $\nu$ of the storage nodes. In addition, some of these storage nodes can also be read by an adversary (eavesdropper). The adversary can access only a limited number of these storage nodes (due to time, memory, geography or other constraints). The user wishes to store information in such a way that the adversary can obtain absolutely no information about the secret message even if he can read any $\mu$ of the storage nodes.

The outline of the paper is as follows. In section \ref{sec:problem_statement} we formally state our problem and give an overview of our notation. In the section \ref{sec:Ozarow-Wyner}, we state the Ozarow-Wyner solution for the wire-tap channel II and perform an analysis using dimension/length profile (DLP) of linear block codes. In section \ref{sec:nested_codes} we present a way to use nested codes, i.e. a set of two codes where one of the codes is a subcode of the other,  on our modified wire-tap channel II and analyze the performance using the DLP of the underlying codes. 

\section{Problem Statement and Formulation}
\label{sec:problem_statement}
\begin{figure}
	\centering
		\includegraphics{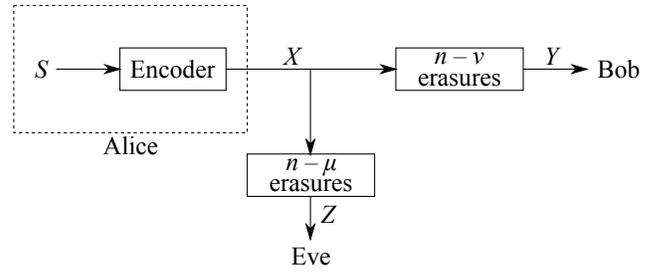}
		\caption{Wiretap Channel II with erasures on the main channel}
		\label{fig:model_overview}
\end{figure}

Let $\mathbb{F}$ be a finite field of size $q$. All the vectors in this discussion will be drawn from vector spaces on $\mathbb{F}$ and the logarithms are taken to the base $q$. The channel under consideration is depicted in fig. \ref{fig:model_overview}. Alice has a uniformly distributed $k$-symbol random message $S$ that must be conveyed to Bob by transmitting a $n$-symbol vector $X$ over the channel. The main channel is such that a fixed number, $n - \nu$, of erasures occur in Bob's received codeword, $Y$. The positions of these erasures are randomly chosen. In addition, there is an eavesdropper who has the ability to tap into any $\mu$ of the $n$ transmitted symbols. Alice knows the values of $\nu$ and $\mu$. She does not know anything else about the erasures on the main channel or the symbols being tapped by Eve. Her task is to choose an encoding scheme which ensures that Bob can completely decode the message while Eve has complete equivocation over the message in spite of knowing the encoding procedure and the $\mu$ symbols revealed to her. A special case of the above problem when $\nu = n$ is the case considered by Ozarow-Wyner \cite{wiretap2}.

First, we define some notation for the analysis of conditional entropy under erasure. Let $I = \{1, 2, 3, \ldots, n\}$ be the index set for the elements in an $n$-symbol vector. Let $J \subseteq I$ be the index set of the revealed positions. Given a vector $X \in \mathbb{F}^n$, by $X_J$, we denote the vector in $\mathbb{F}^{|J|}$ which is formed by taking the elements of $X$ indexed by the elements of the set $J$. In particular, note that $(X_J, J)$ completely describes the result of erasing the symbols of $X$ in positions $I\backslash J$.

Let $M \subset I, W \subset I$ be the index set of the revealed symbols of the main channel and the wire-tapper's channel respectively. Our objective is to devise a coding scheme such that,
\begin{align}
H(S | X_M) & = 0,   & \qquad \forall M \subset I, |M| = \nu \label{eqn:reliability}\\
H(S | X_W) & = H(S),& \quad \forall W \subset I, |W| = \mu \label{eqn:security}
\end{align}
where all the entropies are computed using base-$q$ logarithms. The above conditions can be met only if $\mu \leq \nu - k$. This is because the following necessary conditions must hold.

\begin{enumerate}
\item for $\mu \geq \nu$, and $M \subset W$, $X \to X_W \to X_M$ is a Markov chain and by data processing inequality
    \begin{displaymath}
        H(S|X_M) \geq H(S|X_W)
    \end{displaymath}
Hence, conditions (\ref{eqn:reliability}) and (\ref{eqn:security}) don't hold for this case.
\item for $\mu \leq \nu$ and $W \subset M$, we have the following necessary condition for achieving (\ref{eqn:reliability}) and (\ref{eqn:security})
\begin{eqnarray*}
  k                       & =    & H(S|X_W) - H(S | X_M)  \\
                          & =    & H(S|X_W) - H(S|X_W, X_{M\backslash W}) \\
                          & =    & I(S;X_{M\backslash W} | X_W)                     \\
                          & \leq & H(X_{M\backslash W}|X_W)                             \\
                          & \leq & H(X_{M\backslash W})                                     \\
                          & \leq & \nu - \mu
\end{eqnarray*}

\end{enumerate}

\section{Ozarow-Wyner Coding for Wire-tap II}
\label{sec:Ozarow-Wyner}
In Ozarow-Wyner coding for the wire-tap channel II with a perfect main channel, a $(n, n-k)$ linear code $C^*$ is chosen. This code has $q^k$ cosets and we can construct an arbitrary bijection between the set of all cosets and the set of all possible $k$-symbol messages. For a given message $S$, Alice chooses a random vector from the corresponding coset with uniform probability and transmits it.

Since the main channel is perfect and the encoding is done such that any given n-tuple maps to a unique message, the decoding across the main channel is error-free. So, we have
\begin{displaymath}
H(S|Y) = H(S|X) = 0
\end{displaymath}

Since $X$ is uniformly distributed in $\mathbb{F}^n$, we have $q^{n - |W|}$ possible values (balls) for $X$ with equal likelihood given Eve's observation $X_W$. If we bin these values based on their cosets, the non-empty bins correspond to the possible messages with non-zero probability. The a-posteriori probability of a message is equal to the fraction of balls present in the corresponding bin. It can be shown that the non-empty bins will have the same cardinality, which means that the possible messages are all equally likely.

In the following, we cast the results of \cite{wiretap2} and \cite{133259} in terms of the above \emph{balls and bins} approach and then analyse the conditional entropy of the secret message using DLP properties. This gives a slightly different interpretation of the problem of coset coding with erasures, which is later used again in the next section to analyze the performance of coset coding on the modified wiretap II channel with erasures on the main channel.
 
\subsection{Coset binning}
Let $a \in \mathbb{F}^n$ be a match for $X_W$ in the symbol positions $W$. Let $c$ be any codeword in $C^*$ with $c_W = 0$. Clearly, $c + a$ is also a match for the observation. Let $C^*_{I\backslash W} \triangleq \{c: c \in C^*, c_W = 0\}$. It can be seen that $C^*_{I\backslash W}$ is a subcode of $C^*$. The set $a + C^*_{I\backslash W}$ is the set of all possible matches for $X_W$ in the coset to which $a$ belongs. Hence, the number of balls in a non-empty bin is $|C^*_{I\backslash W}|$. $q^{n - |W|}$ balls distributed in such a fashion will result in $q^{n - |W|}/|C^*_{I\backslash W}|$ non-empty bins. Hence,

\begin{equation}
H(S|X_W) = n - |W| - \mathrm{dim}(C^*_{I\backslash W})
\end{equation}

For complete secrecy, we must achieve (\ref{eqn:security}) or equivalently
\begin{eqnarray*}
min \{ H(S | X_W) : W \subset I, |W| = \mu \}  & = & k \\
\Rightarrow min \{n - |W| - \mathrm{dim}(C^*_{I\backslash W}) : W \subset I, |W| = \mu\} & = & k \\
\Rightarrow n - \mu - max\{\mathrm{dim}(C^*_{I\backslash  W}) : W \subset I, |W| = \mu\} & = & k \\
\Rightarrow n - \mu - k_{n - \mu} (C^*) & = & k
\end{eqnarray*}

Here, $k_i(C^*)$ denotes the $i^{\mathrm{th}}$ dimension/length profile (DLP) of the linear code $C^*$. For a detailed discussion on DLP, see \cite{340452}.

\section{Nested Codes}
\label{sec:nested_codes}
In this section, we consider the case of the modified wiretap II channel where Bob gets only $\nu$ of the $n$ symbols in Alice's transmitted codeword and Eve gets $\mu$ of the codeword symbols. We propose and analyze a coding scheme for this channel. We then show that if the codes are MDS, then the coding scheme achieves the secrecy capacity of this channel.

Let $C$ be an $(n, k)$ code and $C^*$ be an $(n, k^*)$ code, with $0 \leq k \leq n$ and $0 \leq k^* \leq n - k$. Also assume that $C \cap C^* = \{0\}$ and $D = C + C^*$. Let $G$ and $G^*$ be the generator matrices of the codes $C, C^*$ respectively. Let $S$ be the uniformly distributed $k$ symbol secret message, and $E$ be a uniformly distributed random vector of length $k^*$. We transmit $X = SG + EG^*$ and we try to analyze the case when we reveal only a certain number of symbols from $X$. Note that when $k^* = n - k$, this nested coding scheme is the same as the Ozarow-Wyner scheme for Wire-tap II.

The set of valid values of $X$ can be binned into $q^k$ distinct cosets of $C^*$. Every such coset of $C^*$ maps to a distinct message in $\mathbb{F}^k$. Let $J \subset I$ be the index set of the revealed symbols. Given the observation $X_J$, we use the binning approach in the previous section to find the wire-tapper's equivocation. Given $X_J$, the number of possible solutions for $X$ is $|D_{I\backslash J}|$. The size of the non-empty bins is the same as in the previous section. We have,

\begin{eqnarray}
H(S|X_J) & = & \mathrm{dim}(D_{I\backslash J}) - \mathrm{dim}(C^*_{I\backslash J})
\end{eqnarray}

To satisfy the conditions in (\ref{eqn:reliability}) and (\ref{eqn:security}), we must have 
\begin{eqnarray*}
\mathrm{dim}(D_{I\backslash M}) - \mathrm{dim}(C^*_{I\backslash M}) & = & 0 \qquad \forall M \subset I, |M| = \nu\\
\mathrm{dim}(D_{I\backslash W}) - \mathrm{dim}(C^*_{I\backslash W}) & = & k \qquad \forall W \subset I, |W| = \mu
\end{eqnarray*}

\subsection{Using nested MDS codes}
\begin{figure}
	\centering
		\includegraphics[width=0.45\textwidth]{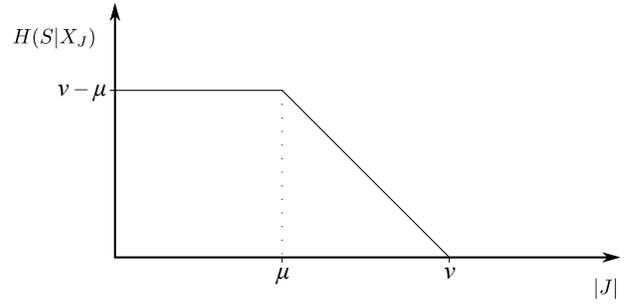}
		\caption{Amount of leaked information vs. number of revealed symbols for the case of nested MDS codes}
		\label{fig:info_leak_graph}
\end{figure}

For a $(n, k)$ maximum distance separable (MDS) code $\widetilde{C}$, we have 
\begin{equation}
\mathrm{dim}(\widetilde{C}_{I\backslash J}) = \max\{0, k - |J|\}
\end{equation}
Hence, if we choose the codes $D$ and $C^*$ to be nested MDS codes, we will have
\begin{align}
\mathrm{dim}& (D_{I\backslash J}) - \mathrm{dim}(C^*_{I\backslash J}) \\
										& = \max\{0, k + k^* - |J|\} - \max\{0, k^* - |J|\} \\
										& = \left\{	\begin{array}{rl}
																								0, 								& |J| \geq k + k^* \\
																								k + k^* - |J|,			& k^* \leq |J| < k + k^* \\
																								k, 								& 0 \leq |J| < k^*
																\end{array}
												\right.
\end{align}
A sketch of the plot of the above function vs. $|J|$ is shown in fig \ref{fig:info_leak_graph} for the case when $k = \nu -\mu$, $k^* = \mu$. 
 	
Suppose there is a situation where $n, \mu, \nu$ are specified and we are free to choose the symbol alphabet $\mathbb{F}$. From the analysis in the previous section, we can achieve the maximum possible secret information rate by choosing $\mathbb{F}$ to be a field of size not less than $n$ and construct two nested Reed-Solomon codes $D, C^*$ of dimensions $\nu, \mu$ respectively. We then have,
\begin{align}
\mathrm{dim}(D_{I\backslash M}) - \mathrm{dim}(C^*_{I\backslash M}) & = 0,& \forall M \subset I, |M| = \nu \\
\mathrm{dim}(D_{I\backslash W}) - \mathrm{dim}(C^*_{I\backslash W}) & = \nu - \mu,& \forall W \subset I, |W| = \mu
\end{align}

In this case, we have a coding scheme that achieves the maximum possible secret information rate. Table \ref{table:rs_example} illustrates how to choose nested MDS codes for the erasure-erasure wiretap channel.

\begin{table}
\centering
\begin{tabular}{|l|p{2in}|}
\hline
Channel parameters & $n = 255$, $\nu = 200$, $\mu = 150$ \\
\hline
Secrecy capacity & $\frac{50}{255} \approx 0.196$ \\
\hline
Code $D$ & $(255, 200)$ RS code over $F_{256}$. Generator polynomial $g(x) = (x - \alpha)(x - \alpha^2)(x - \alpha^3) \cdots (x - \alpha^{55})$ \\
\hline
Code $C^*$ & $(255, 150)$ RS code over $F_{256}$. Generator polynomial $g(x) = (x - \alpha)(x - \alpha^2)(x - \alpha^3) \cdots (x - \alpha^{105})$ \\
\hline
\end{tabular}
\caption{Nested MDS coding scheme example}
\label{table:rs_example}
\end{table}

\section{Conclusion}
In this paper, we have studied the erasure-erasure wiretap channel model where the numbers of erasures are fixed but the positions of the erasures are chosen at random. We have shown that a coding scheme based on nested MDS codes achieves the secrecy capacity of this channel. We have assumed that the channel model permits us to choose the finite field over which we draw the symbols. Analyzing the secret information rate of general (non-MDS) nested codes over a similar channel is a natural generalization of our problem. This will also lead us to design secure coding schemes for a much wider choice of channels.

\bibliographystyle{IEEEtran}
\bibliography{coset_coding_isit}

\begin{thebibliography}{1}
\providecommand{\url}[1]{#1}
\csname url@samestyle\endcsname
\providecommand{\newblock}{\relax}
\providecommand{\bibinfo}[2]{#2}
\providecommand{\BIBentrySTDinterwordspacing}{\spaceskip=0pt\relax}
\providecommand{\BIBentryALTinterwordstretchfactor}{4}
\providecommand{\BIBentryALTinterwordspacing}{\spaceskip=\fontdimen2\font plus
\BIBentryALTinterwordstretchfactor\fontdimen3\font minus
  \fontdimen4\font\relax}
\providecommand{\BIBforeignlanguage}[2]{{%
\expandafter\ifx\csname l@#1\endcsname\relax
\typeout{** WARNING: IEEEtran.bst: No hyphenation pattern has been}%
\typeout{** loaded for the language `#1'. Using the pattern for}%
\typeout{** the default language instead.}%
\else
\language=\csname l@#1\endcsname
\fi
#2}}
\providecommand{\BIBdecl}{\relax}
\BIBdecl

\bibitem{wiretap1}
A.~D. Wyner, ``The wire-tap channel,'' \emph{Bell Syst. Tech. J.}, vol.~54,
  no.~8, pp. 1355--1387, oct 1975.

\bibitem{1055892}
I.~Csiszar and J.~Korner, ``Broadcast channels with confidential messages,''
  \emph{Information Theory, IEEE Transactions on}, vol.~24, no.~3, pp.
  339--348, May 1978.

\bibitem{wiretap2}
L.~H. Ozarow and A.~D. Wyner, ``Wire-tap channel {II},'' \emph{Bell Labs Tech.
  J.}, vol.~63, no.~10, pp. 2135--2157, dec 1984.

\bibitem{1003821}
R.~Zamir, S.~Shamai, and U.~Erez, ``Nested linear/lattice codes for structured
  multiterminal binning,'' \emph{Information Theory, IEEE Transactions on},
  vol.~48, no.~6, pp. 1250--1276, Jun 2002.

\bibitem{359176}
A.~Shamir, ``How to share a secret,'' \emph{Commun. ACM}, vol.~22, no.~11, pp.
  612--613, 1979.

\bibitem{133259}
V.~Wei, ``Generalized hamming weights for linear codes,'' \emph{Information
  Theory, IEEE Transactions on}, vol.~37, no.~5, pp. 1412--1418, Sep 1991.

\bibitem{340452}
J.~Forney, G.D., ``Dimension/length profiles and trellis complexity of linear
  block codes,'' \emph{Information Theory, IEEE Transactions on}, vol.~40,
  no.~6, pp. 1741--1752, Nov 1994.

\bibitem{1397965}
Y.~Luo, C.~Mitrpant, A.~Vinck, and K.~Chen, ``Some new characters on the
  wire-tap channel of type ii,'' \emph{Information Theory, IEEE Transactions
  on}, vol.~51, no.~3, pp. 1222--1229, March 2005.

\end{thebibliography}

\end{document}